\documentclass[onecolumn,12pt,preprintnumbers,amsmath,amssymb]{revtex4}

\usepackage{epsf}
\usepackage{graphicx}  
\usepackage{dcolumn}   
\usepackage{bm}        

\newcommand{\be}{\begin{equation}}
\newcommand{\en}{\end{equation}}
\newcommand{\ba}{\begin{array}}
\newcommand{\ea}{\end{array}}
\newcommand{\bea}{\begin{eqnarray}}

\newcommand{\ena}{\end{eqnarray}}

\begin{document}

\preprint{BUN/0312-2007}

\title{Natural Phantom Dark Energy, Wiggling
Hubble Parameter $H(z)$ and Direct $H(z)$ Data
 }

\author{Hongsheng Zhang and
  Zong-Hong Zhu\footnote{E-mail address: zhuzh@bnu.edu.cn}}
\affiliation{
 Department of Astronomy, Beijing Normal University, Beijing 100875, China}


\begin{abstract}
 Recent direct $H(z)$ data indicate that the parameter
 $H(z)$ may wiggle with respect to
 $z$. On the other hand the luminosity distance data of supernovae
 flatten the wiggles of $H(z)$ because of integration effect. It is
 expected that the fitting results can be very different in a model
 permitting a wiggling $H(z)$ because the data of supernovae is
 highly degenerated to such a model. As an example the natural phantom dark energy
  is investigated in this paper. The dynamical property of this model is studied.
  The model is fitted by the direct $H(z)$ data set and the SNLS
  data set, respectively. And the results are quite
  different, as expected. The quantum stability of this model is also
  shortly discussed. We find it is a viable model if we treat it as an
  effective theory truncated by an upperbound.

\end{abstract}


 \maketitle

\section{Introduction}

 The acceleration of the universe is one of the most significant cosmological discoveries
 over the last decades \cite{acce}. The decisive evidence of the present
 acceleration is witnessed by supernovae. The principle of this conclusion
 is based on the fitting of LCDM (cold dark matter
 with a cosmological constant) model by the data of the luminosity
 distance of the supernovae. The phenomena of this cosmological
 acceleration is of very interest, for which a large number of models have been
 proposed besides LCDM, and several of them have been
 fitted by luminosity distances of supernovae, for a review, see
 \cite{review}.

  In a cosmological model, one calculates the
 luminosity distances as follow,
 \be
 D_l=\frac{1+z}{H_0}\int_0^z\frac{H_0}{H(z')}dz',
 \en
 where $H(z')$ denotes the Hubble parameter, and $H_0$ represents its present value.
 Then one can constrain
 the parameters in the model by using the observation data of supernovae through
 $\chi^2$ or other method. A deficiency of this method is that one
 obtains luminosity distance through an integrating to Hubble parameter
 $H$, and therefore, the fine structures, such as wiggles on $H$,
  can not show themselves in such a method.
 For example, compared with
 \be
 H(z)/H_0=1,
 \en
 \be
 H(z)/H_0=\frac{1}{1+\sin(nz)},
 \en
 is surely a different model. But for a large $n$, they always share the same
  confidence region in fittings by using luminosity distances data.
  And different $n$ (for large $n$) also share the same confidence
  region. We see that some types of fine structures of $H$ is highly
  degenerate to the luminosity distance data. To break this
  degeneration one needs the observational data of $H(z)$, not only an
  integration of $H^{-1}(z)$.

   Fortunately there is a newly
 developed scheme to obtain the Hubble parameter directly
 at different redshift \cite{h(z)}, which is based on a method
 to estimate the differential ages of the oldest galaxies
 \cite{age}. By using of the previously released data \cite{aid},
 Simon {\it et al.} obtained a sample of direct $H(z)$ data in the interval $z\in
 (0,1.8)$~\cite{simon}, almost as the same interval of the data of
 luminosity distances from supernovae. We show this sample in table
 I.

\begin{table}[]
\begin{center}
\begin{tabular}{c|lllllllll}\hline
 $z$ &\ 0.09 & 0.17 & 0.27 & 0.40 & 0.88 & 1.30 & 1.43
 & 1.53 & 1.75\\ \hline
 $H(z)\ ({\rm km~s^{-1}\,Mpc^{-1})}$ &\ 69 & 83 & 70
 & 87 & 117 & 168 & 177 & 140 & 202\\ \hline
 68.3\% confidence interval &\ $\pm 12$ & $\pm 8.3$ & $\pm 14$
 & $\pm 17.4$ & $\pm 23.4$ & $\pm 13.4$ & $\pm 14.2$
 & $\pm 14$ &  $\pm 40.4$\\ \hline
\end{tabular}
\end{center}
\caption{\label{tabhz} The direct observation data of $H(z)$
 ~\cite{simon}
 (see~\cite{ratra} also).}
\end{table}
   Table I displays an unexpected feature of $H(z)$: It decreases
   with respect to the redshift $z$ at redshift $z\sim 0.3$
   and $z\sim 1.5$, which means that the total
   fluid in the universe behave as phantom. This information of
    dynamical property
   of the universe is very difficult to be drawn from the data of
   supernovae. This feature of $H(z)$ implicates that the dark
   energy  component of the cosmic fluid must behave as phantom sometime, which can be proved
   by the following argument.
 In standard general relativity for a spatially flat universe, which is implied
 either by theoretical side (inflation in the early universe)
 ,or observation side (CMB fluctuations \cite{WMAP}), the Friedmann
 equation reads,
 \be
 H^2=\frac{1}{3\mu^2}(\rho_m+\rho_{de}),
 \label{fried}
  \en
  where $\rho_m$ denotes the density of dust matter, $\rho_{de}$
  stands for the density of dark energy, and $\mu$ represents the
  reduced Planck mass. Differentiate with respect
  to the redshift $z$, we derive from (\ref{fried}),
  \be
  2H\frac{dH}{dz}=\frac{1}{3\mu^2}({\frac{d\rho_m}{dz}}+\frac{d\rho_{de}}{dz}),
  \en
  where a prime denotes derivation with respect to $z$. Clearly,
  if ${dH}/{dz}<0$ at some redshift (as shown in table I), one concludes ${d\rho_{de}}/{dz}<0$
  since ${d\rho_m}/{dz}>0$, which means the dark energy behaves as
  phantom.

     The present (or at very low redshift) phantom behavior of dark energy
     is also implied by the supernovae data~\cite{call}. Generally speaking, a simple
      phantom field (scalar field with kinetic term of false sign)
     is quantum mechanically unstable. However,
     several evidences imply that our present 4 dimensional standard model and
     general relativity is not the final theory. The phantom model
     can be treated as reduced theory of more fundamental theory, in which there is
     no field behaves as phantom \cite{review}. Thus the stability problem may be evaded.  Actually,
      many of the reduced theories do contain
    phantoms, as the ones coming from string and/or M-theory
   compactification, or higher-derivative supergravities, or
  modifications of Einstein gravity itself, for example,
  such a field may be motivated from S-brane constructions
  in string theory \cite{sbrane}. Moreover, there exist examples in which
   an effective phantom and/or quintessence description
 of the late time universe naturally emerges, even when the starting
 theory does not clearly show the phantom and/or quintessence
 structure \cite{eno}. Therefore it may be reasonable to
 investigate such models as an effective theory. Phenomenologically, the cosmological
  models with phantom
     matter have been investigated
  extensively \cite{phantom}.   Also, urged by observations, the
   models with dark energy whose
   EOS crosses $-1$ have been investigated in \cite{cross}.

   However, the $H(z)$ data in table I implies the EOS of $total~ fluid$ in the
   universe crosses $-1$, not only the dark energy sector.
   Moreover, the phase oscillation over deceleration phase and
   acceleration phase is clear through the history of the universe.
   In a fitting in frame of LCDM model, the point $z\sim 1.5$ is
   determinately beyond 1-$\sigma$ level \cite{simon}. Furthermore, it is shown that the data
   point near $z\sim 1.5$, which dips so sharply and stays clearly outside of
   the best-fit of the LCDM, XCDM and $\phi$CDM models studied in
   \cite{ratra}. Contrarily, a study show that the model whose Hubble parameter
   is directly endowed with oscillating ansatz by parameterizations
   fit the data much better than those of LCDM, IntLCDM, XCDM, IntXCDM,
   VecDE, IntVecDE \cite{weihao}. However no previous physical dark energy models possessing this
   oscillating property. Therefore, it deserves to present a physical model in which the EOS of total~fluid
   crosses $-1$.

  In this paper we put forward a model in which phantom field with natural potential
 ,ie, the potential of a pseudo Nambu-Goldstone Boson (PNGB), drives the universe. We
  shall show that in such a model all the features of $H$ in table I
  can be realized naturally, and the fitting results of the parameters in this model
  are rather different according to supernovae and direct $H(z)$
  data. PNGB is an important idea in particle physics. It emerges
  whenever a global symmetry is spontaneously broken. There are two
  key scales of PNGB generation. One is the scale at which the global
  symmetry breaks, denoted by $f$, and the other is the scale at
  which the soft explicit symmetry breaks, denoted by $C$. Under
  this assumption the potential of PNGB reads,
  \be
  V=C^4\left(1\pm \cos(\frac{N\phi}{f})\right).
  \label{natural}
  \en
  Inflation model driven by a scalar with such a potential was
  firstly studied in \cite{natural inflation}. Generally speaking in the context of inflation
  model the cosine function in potential never completes a cycle.
  The scalar PNGB can also play the role of dark energy \cite{PNGB dark energy}.
  In this scenario the energy scale of the  the global
  symmetry breaking $f$ keeps about the same as the case of inflation model
  ,ie, the Planck scale. Contrarily, the scale of explicit symmetry breaking
  decreases to an extremely low scale, ie, $10^{-3}$eV, which is comparable to
  neutrino mass yielded by Mikheyev-Smirnov-Wolfenstein (MSW)
  mechanism. Phenomenologically, the natural potential has been
  generated to solve the coincidence problem, in which the cosine function
  in potential oscillates many cycles \cite{coin}, and therefore the
  densities of dark energy and dust can be comparable several times in
  the history of the universe. But the previous models with PNGB dark
  energy can not realize the feature that the EOS of $total~ fluid$ in the
   universe crossing $-1$. This feature appears naturally in the
   present phantom natural dark energy model.

  In the next section we shall present the phantom natural dark energy model and
  investigate some dynamical properties of it. In section III, we fit
  this model by using the SNLS  data and direct $H(z)$ data, respectively. The
  main conclusions and some discussions appears in the last section.

\section{The Model}
 We work in the frame of the standard 4 dimensional general
 relativity. The phantom is characterized by a false sign of kinetic term in
 the Lagrangian,
 \be
 {\mathcal L_p}=\frac{1}{2}\partial_{\mu} \phi \partial^{\mu} \phi
 -V(\phi),
 \en
 where and in the follow, we take the signature $(-,+,+,+)$.
   In the present model a phantom field with generalized
 natural potential plays the role of dark energy.
 In an FRW universe, $\rho_{de}$ in (\ref{fried}) becomes
  \be
  \rho_{\phi}=-\frac{1}{2}\dot{\phi}^2+V(\phi),
  \en
  and the pressure of the scalar reads
  \be
  p_{\phi}=-\frac{1}{2}\dot{\phi}^2-V(\phi),
  \en
  where a dot denotes derivative with respect to time. And the equation of
  motion of $\phi$ reads,
  \be
  \label{eom}
  -\ddot{\phi}+3H\dot{\phi}+\frac{dV}{d\phi}=0.
   \en
   Based on the
  former researches, we phenomenologically generalize the natural potential to the
  following form,
  \be
  V(\phi)=V_0\left(1+A\cos(p\frac{\phi}{\mu})\right).
  \label{potential}
  \en
   With the new
  dimensionless variables below,
  \bea
  x&\triangleq&\frac{\dot{\phi}}{\sqrt{6}\mu H},\\
  y&\triangleq&\frac{\sqrt{V}}{\sqrt{3}\mu H},\\
  l&\triangleq&\frac{\sqrt{\rho_m}}{\sqrt{3}\mu H},\\
  b&\triangleq&\frac{\sqrt{V_0}}{\sqrt{3}\mu H},
  \ena
  the dynamics of the universe can be described by the following
  dynamical system,
 \bea
 \label{1}
 x'&=&\frac{3}{2}
 x(-2x^2+l^2)-3x+\frac{\sqrt{6}}{2}p\sqrt{A^2b^4-(y^2-b^2)^2},\\
 \label{2}
  y'&=&\frac{3}{2}
 y(-2x^2+l^2)-\frac{\sqrt{6}}{2}p xy^{-1}\sqrt{A^2b^4-(y^2-b^2)^2},\\
 \label{3}
   l'&=&\frac{3}{2}
 l(-2x^2+l^2)-\frac{3}{2}l,\\
 \label{4}
  b'&=&\frac{3}{2}
 b(-2x^2+l^2),
 \ena
 where
  a prime stands for derivation with respect to
 $s\triangleq\ln(1+z)$.
  Note that the 4 equations (\ref{1}), (\ref{2}),
 (\ref{3}), (\ref{4}) of this system are not independent. By using the Friedmann
 constraint, which can be derived from the Friedmann equation,
 \be
 \label{constraint}
 -x^2+y^2+l^2=1,
 \en
 the number of the independent equations can be reduced to 3.
    There
 are
 four critical points of this system satisfying $x'=y'=l'=b'=0$ appearing at
 \bea
 \label{sin1}
 &x&=l=0,~~y=1, ~~b=\pm \sqrt{\frac{1}{1+A}}~;\\
 &x&=l=0,~~y=1, ~~b=\pm \sqrt{\frac{1}{1-A}}~.
 \label{sin2}
  \ena
 All of them satisfy the Friedmann constraint
 (\ref{constraint}). To obtain real values of the variables at the singularities
 we see that if $A\geq 1$ only the former two
 exist, if $A\leq -1$ only the latter two exist, and only for
 $-1<A<1$ all of the four critical points exist. The critical points imply that the universe will enter a
 pure dark energy phase at last, if the singularity is stationary.
   To investigate the properties of the dynamical system in the
  neighbourhood of the singularities, impose a perturbation
  to the critical points,
 \bea
  \delta x'&=&E_{11}\delta x+E_{12}\delta y+E_{14}\delta b,
  \label{linear1}
  \\
  \delta y'&=&E_{22}\delta y+E_{24} \delta b,
  \label{linear2}
  \\
  \delta l'&=&E_{33}\delta l,
  \label{linear3}
  \\
  \delta b'&=&0,
  \label{linear4}
  \ena
  where we have used (\ref{sin1}) or (\ref{sin2}), and the components of
  the eigenmatrix reads,
  \bea
  E_{11}&=&-3,\\
  E_{12}&=&-\sqrt{6}py(y^2-b^2)\alpha ^{-1},\\
  E_{14}&=&\sqrt{6}p[A^2b^3+(y^2-b^2)b]\alpha ^{-1},\\
  E_{22}&=&\sqrt{6}xp(y^2-b^2)\alpha ^{-1},\\
  E_{24}&=&-\sqrt{6}pxy^{-1}[A^2b^3+b(y^2-b^2)]\alpha^{-1},\\
  E_{33}&=&-3/2,
  \ena
  where
  \be
  \alpha\triangleq\sqrt{A^2b^4-(y^2-b^2)^2}~.
  \en
  The 4 eigenvalues of this linear system reads
  \be
  \lambda_{1}=-3,~\lambda_{2}=\sqrt{6}xp(y^2-b^2)\alpha ^{-1},~
  \lambda_{3}=-3/2,~\lambda_{4}=0.
  \en
  The property of $\lambda_2$ is rather complicate around the
  singularities. For example, it goes to different values along different pathes
  around the singularity $x=0,~z=0,~b=1/\sqrt{1+A}~$.
  The 6 repeated limits read,
  \bea
  \lim_{z\to 0}\lim_{x\to 0}\lim_{b\to \frac{1}{\sqrt{1+A}}}\lambda_2&=&0,\\
  \lim_{x\to 0}\lim_{z\to
  0}\lim_{b\to\frac{1}{\sqrt{1+A}}}\lambda_2&=&ip\sqrt{\frac{3A}{1+A}}~,\\
  \lim_{z\to 0}\lim_{b\to \frac{1}{\sqrt{1+A}}}\lim_{x\to
  0}\lambda_2&=&0,\\
  \lim_{b\to \frac{1}{\sqrt{1+A}}}\lim_{z\to 0}\lim_{x\to
  0}\lambda_2&=&0,\\
  \lim_{b\to \frac{1}{\sqrt{1+A}}}\lim_{x\to
  0}\lim_{z\to 0}\lambda_2&=&0,\\
  \lim_{x\to 0}\lim_{b\to \frac{1}{\sqrt{1+A}}}\lim_{z\to
  0}\lambda_2&=&ip\sqrt{\frac{3A}{1+A}}~.
  \ena

  Hence the limit of $\lambda_2$ does not exist at
  the singularities. However, we see that the real parts of the
  limits
  keep zero independent of pathes, which means that the system reaches
  an indifferent equilibrium. In such a de Sitter universe at the critical
  point the kinetic energy of the phantom and dust matter vanish,
  but the potential energy can reside at any values, which depends on the initial values
  of kinetic energy, potential energy, dust density and the Hubble
  parameter.

 As we have pointed out in section I, the data of supernovae
 is insensitive to the oscillating behaviour of $H(z)$. In this
 section we show the fitting results by the direct $H(z)$ data and
 SNLS data by $\chi^2$-statistics, respectively. The $H(z)$ data have been used to
 constraint models in \cite{ratra} \cite{weihao} \cite{zyi}. Here we adopt SNLS data \cite{snls},
 which is believed to be more consistent with CMB data. Figure 2 displays the
 fitting results. We set $A=1$, which means we adopt the original PNGB potential, $b(z=0)=0.616$, $\phi/\mu
 (z=0)=0.022$, $H_0=72{\rm km~s^{-1}\,Mpc^{-1}}$ \cite{free}.
  In figure 2 we find an extraordinary property of
 (a): the 68.3\% confidence contour is disconnect. The physical
 explanation is that the data set of direct $H(z)$ is too small, that is,
 the data do not distinctly illuminate how many ``wiggles" inhabit on $H(z)$. New wiggles may hide
 in the gaps of the data set, which leads that a much bigger $p$
 lies in the same confidence region as a smaller $p$. (b) clearly
 shows that the resolution of supernavae data is very inefficiency to
 the oscillating behaviour of $H(z)$.
  \begin{figure}
 \centering
 \includegraphics[totalheight=2.7in, angle=0]{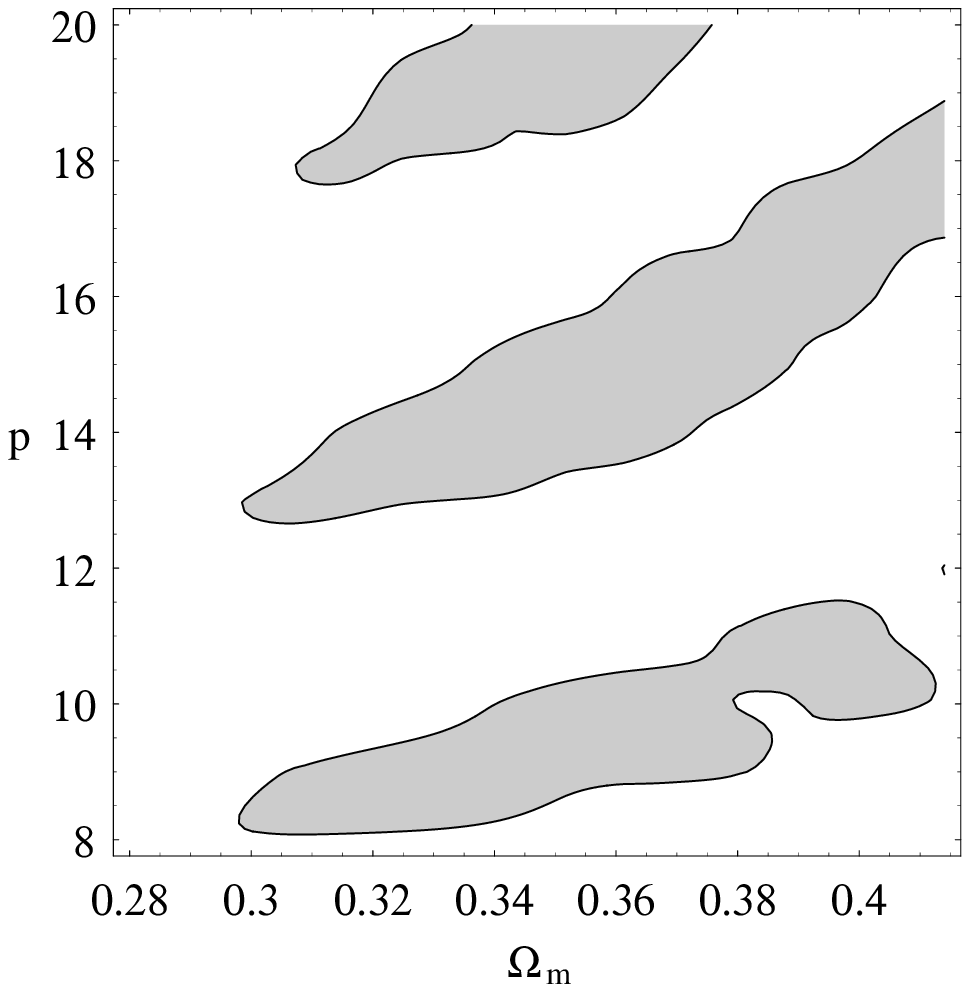}
 \includegraphics[totalheight=2.7in, angle=0]{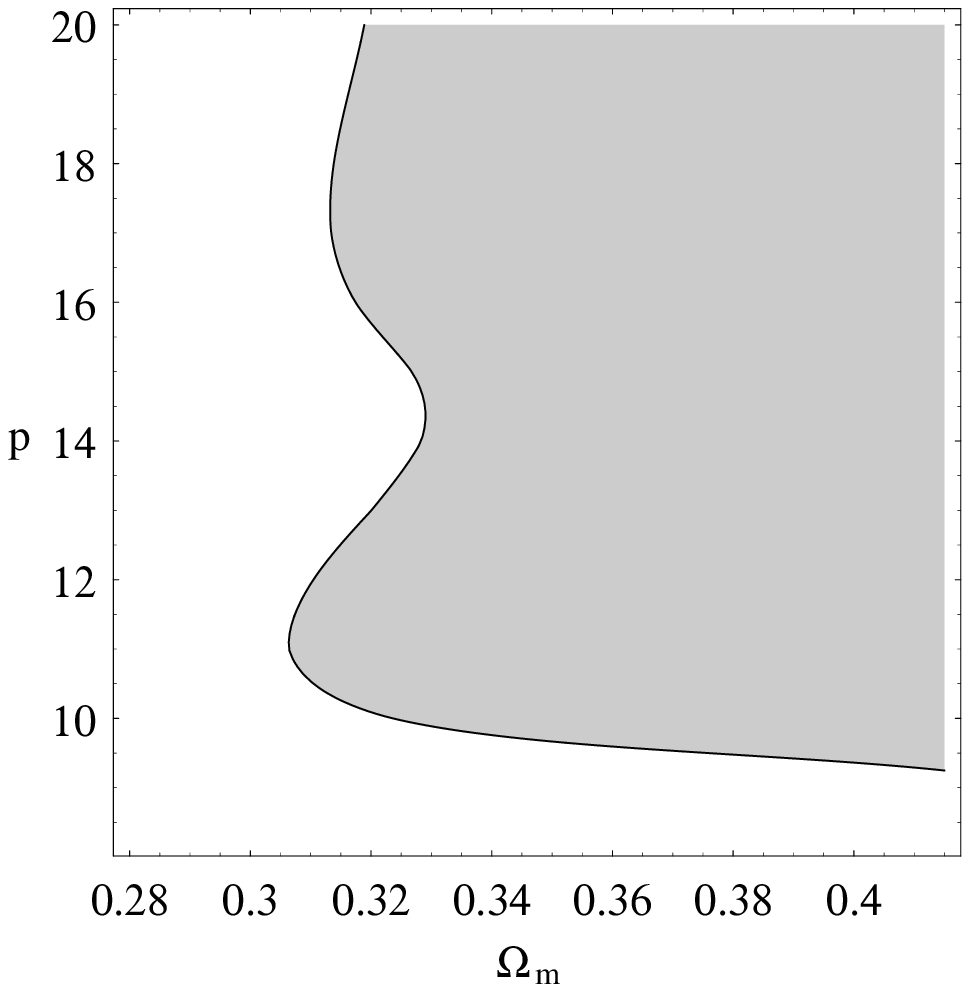}
 \caption{The fitting result of the parameters $\Omega_m$ ($l_0^2$)
 and $p$. {\bf{(a)}} The 68.3\% confidence contour plot  by using the
 direct $H(z)$ data in table I. {\bf{(b)}} The 68.3\% confidence
 contour plot by using the SNLS data.}
 \label{fitting}
 \end{figure}
 We show the deceleration parameter $q$ in figure 2 with best fit
 values of $\Omega_m$ and $p$ by direct $H(z)$ data. Figure 2
 illuminates that the universe oscillates between deceleration phase
 and acceleration phase.
 \begin{figure}
 \centering
 \includegraphics[totalheight=2.5in, angle=0]{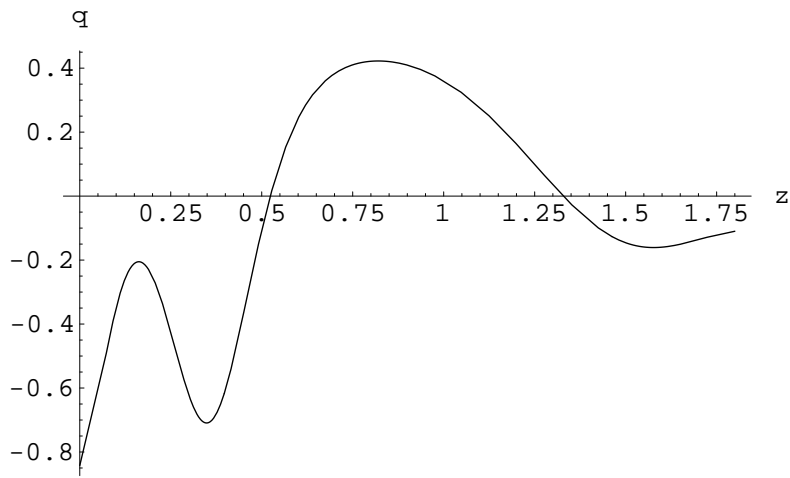}
 \caption{The deceleration parameter $q$ with best fit
 $\Omega_m=0.397$ and $p=11.0$. }
 \label{dece}
 \end{figure}

 \section{Quantum Stability}
 A severe problem of any phantom field is quantum stability. In
 practice, we do not require that the phantom is fundamentally
 stable, but quasi-stable, which means, its lifetime is larger
 than the age of the universe. This problem has been discussed in \cite{carroll} \cite{cline}.
   Here we follow the investigations in
 \cite{carroll}. The simplest interaction between phantom and
 graviton takes the form,
 \be
  \phi \rightarrow h + \phi_1 + \phi_2 \ ,
  \label{intergra}
\en
 where $h$ denotes the gravitational fluctuations on an FRW
 background, $\phi_1$, $\phi_2$ represent other two phantom fields.
 We note here that, though $h$ will be different if we take a
 Minkowski background, but, the difference is tiny and negligible in the spacetime
 region we considered for this interaction. Here we consider a series expansion
 around the initial value of the numerical
 example we studies above, $\phi(z=0)/\mu=0.022$. Based on the discussion in
 \cite{carroll}, we set the interaction term,
 \bea
 {\mathcal L}_{i} &=& \frac{1}{\mu}(\mu h)
        \frac{1}{3!}V^{\prime\prime\prime}(\phi_0)\phi^3 \cr\cr
        &=& A\lambda_e(\mu h)\phi^3\ ,
 \ena
 Where $\lambda_e$ is defined as
 \be
 \lambda_e\triangleq
 \frac{1}{6}\frac{V_0}{\mu^4}\sin(p\frac{\phi_0}{\mu})\sim
 10^{-118},
 \en
 where $p$ takes the best fit value in last section $p=15.3$.
 Clearly, if we treat phantom as fundamental theory, it will be
 unstable, and the reaction rate goes to infinity because the volume
 of the phase space of $\phi$, $\phi_1$, $\phi_2$ goes to infinity.
 However, if we treat it as an effective theory which is only valid
 below some energy scale $\Lambda$, the reaction rate $\Gamma$
 becomes,
\be
   \Gamma \sim \lambda_e^2 \frac{\Lambda^2}{m_{\phi}}\ ,
\en
 where the effective mass of the phantom $m_{\phi}$ is defined as
 \be
 m_{\phi}\triangleq (-AV'')^{1/2}\sim
 p\mu\sqrt{\frac{AV_0}{\mu^4}}=10^{-60}p\mu A^{1/2}.
 \label{A}
 \en
 Here and the following, we take $A=1$ without special announcement.
 The phantom field as an effective field is viable if its reaction
 rate $\Gamma$ is smaller than the present Hubble parameter,
 \be \Gamma<H_0\sim10^{-60}\mu, \en
 which means,
 \be
 \lambda_e^2 \frac{\Lambda^2}{m_{\phi}}<H_0,
 \en
 that is $\Lambda<10^{58}\mu$. In fact, any effective theory is valid
  at such a high energy scale, which is far beyond our present lab energy scale,
  surely be a perfect effective theory.
      But, besides the decay channel as shown in (\ref{intergra}), we must consider the cases
      that one phantom decay into several particles. By summing over
      all these possibilities, one arrives at the total reaction
      rate \cite{carroll}
      \be
 \Gamma_{\rm tot} =
        \Gamma\left[1-\left(\frac{\Lambda}{\mu}\right)^2\right]^{-2} \ .
 \en
  We see that if $\Lambda$ is smaller than $\mu$, $\Gamma_{\rm tot}$
  will keep the same order of $\Gamma$, which is not a stringent
  constraint.

  However, as an effective theory, one must include all possible terms compatible with
   the symmetry of the Lagrangian up to finite orders to guarantee the
   renormalizablity of the theory--contributions
 from high order terms are much suppressed which
  we can neglect up to required precision. A most famous
  effective theory is four-fermion interaction theory, as the low
  effective theory of electro-weak interaction. Expands the
  propagator of the gauge boson in electro-weak according to the
   mass of $W$ boson, we see that the derivative coupling appears.
   Hence a derivative coupling in an effective theory is quite
   reasonable. We
   consider the interaction Lagrangian of graviton and phantom, with an approximate global symmetry,
  \be
 {\cal L}_{ig}= \frac{\gamma}{\mu^2\Lambda^4}  \left[\mu
 h(\partial\phi, \partial\phi) \right]^2 \ ,
 \en
 where $\gamma$ is a constant of order 1. The reaction rate becomes
 \be
  \Gamma \sim {\gamma^2\Lambda^6 \over m_\phi \mu^2}\
  \label{gammagi} ,
\en
 which should be smaller than the present Hubble parameter. Therefore
 we reach
 \be
 \Lambda^6<\frac{H_0m_{\phi}\mu^4}{\gamma^2}\sim 10^{-120}p\mu^6.
 \en
 The key difference between our result and the result in
 \cite{carroll} dwells at the effective mass of the phantom field.
 In fact, our result the reaction rate $\Gamma$ in (\ref{gammagi}) is
  is suppressed by a factor $p$. On the observational side, we see
  that the 1-$\sigma$ confidence region form a confidence tower, no
  clear upper bound of $p$. On the theoretical side, one hardly find
  principles to  constrain $p$ in the natural potential in this
  cosmological context. Physically, the effective mass of the phantom in the present model can be
  notably larger than $H_0=10^{-33}$ eV, which is taken as the mass
  of the phantom in \cite{carroll}. For example, if
  $m_{\phi}=10^{-27}$ eV, which is quite beyond our present abilities of
  accelerators, $\Lambda$ will exceed 1TeV and hence it is also beyond
  our present lab energy scale. Thus, the present model is promising
  due to this suppress mechanism for derivative coupling. Also, we
  note here that a larger $A$ is helpful to increase the mass of the
  phantom, which can be seen from (\ref{A}).

\section{Conclusions and discussions}

 To summarize, this paper illuminates that direct $H(z)$ data is
 much more efficient than the supernovae for the fine structures of
 Hubble diagram.

 We first put forward a model based on the previous studies on the
 PNGB. In this model the total fluid in the universe
 may evolve as phantom in some stages, which contents the direct $H(z)$ data
 in table I.  Then we study its dynamical properties and find its critical
 points. And we also study the stability about the singularities of
 this system.

 In section II we fit our model by using $H(z)$ data and supernovae
 data, respectively. The results are quite different, as we expected.
 Because the sample of $H(z)$ data is too small, the confidence
 contour is disconnect, which means that we  still lack enough information
 about the oscillations of $H(z)$. We hope the future observations
 offering more data of $H(z)$ so that we can investigates the
 history of the universe in a more detail way.

 In section III we investigate the stability of the present model.
 Our treatise is to treat the phantom model as an effective model
 truncated at some energy scale $\Lambda$. As the previous studies,
 we find that the coupling to graviton needs a truncate scale much
 larger than the lab energy scale, if we require the lifetime of the
 phantom is longer than the universe. Different from the previous
 studies, we find that the derivative coupling between phantom and graviton is viable
 due to the special potential of the present model.


{\bf Acknowledgments:}
 We thank M. Trodden for discussions on his paper \cite{carroll}. Our thanks also
 goes to F. Wang for discussions of the section III. This work was supported by
  the National Natural Science Foundation of China
    , under Grant No. 10533010, the Project-sponsored SRF for ROCS, SEM of China,
    and  Program for New Century Excellent Talents in University (NCET).

\end{document}